\documentclass{article}
\usepackage{graphicx}
\usepackage{chicago}
\usepackage[below]{placeins}
\usepackage{amssymb}
\newcommand{\beq}{\begin{equation}}
\newcommand{\eeq}{\end{equation}}
\newcommand{\beqa}{\begin{eqnarray}}
\newcommand{\eeqa}{\end{eqnarray}}
\renewcommand{\(}{\left(}
\renewcommand{\)}{\right)}

\newcommand   {\half}    {\frac{1}{2}}

\newcommand{\AND}{\mbox{\footnotesize\tt AND}}
\newcommand{\OR}{\mbox{\footnotesize\tt OR}}
\newcommand{\sign}{\mbox{sign}}
\newcommand{\s}{{\mathbf s}}
\renewcommand{\v}{{\mathbf v}}
\newcommand{\Mm}{{\cal M}_m}
\begin{document}

\begin{titlepage}

\begin{flushright}
LU TP 00-19 \\
\end{flushright}

\vspace{.25in}

\LARGE

\begin{center}
{\bf An Information-Based Neural Approach to Constraint Satisfaction} \\
\vspace{.3in}
\large

Henrik J\"onsson\footnote{henrik@thep.lu.se} and
Bo S\"oderberg\footnote{bs@thep.lu.se}\\
\vspace{0.10in}
Complex Systems Division, Department of Theoretical Physics\\
Lund University,  S\"olvegatan 14A,  S-223 62 Lund, Sweden \\
{\tt http://www.thep.lu.se/complex/}

\vspace{0.25in}

To appear in {\it Neural Computation}

\end{center}

\large

\vspace{0.6in}

\begin{center}
{\bf Abstract}
\end{center}

A novel artificial neural network approach to constraint satisfaction
problems is presented. Based on information-theoretical
considerations, it differs from a conventional mean-field approach in
the form of the resulting free energy. The method, implemented as an
annealing algorithm, is numerically explored on a testbed of $K$-SAT
problems. The performance shows a dramatic improvement to that of a
conventional mean-field approach, and is comparable to that of a
state-of-the-art dedicated heuristic (Gsat+Walk). The real strength of
the method, however, lies in its generality -- with minor
modifications it is applicable to arbitrary types of discrete
constraint satisfaction problems.

\end{titlepage}

\large

\newpage

\section{Introduction}

In the context of difficult optimization problems, artificial neural
networks ({\em ANN}) based on the mean-field approximation provides a
powerful and versatile alternative to problem-specific heuristic
methods, and have been successfully applied to a number of different
problem types \shortcite{hop85,pet98}.

In this paper, an alternative ANN approach to combinatorial constraint
satisfaction problems ({\em CSP}) is presented. It is derived from a
very general information-theoretical idea, which leads to a modified
cost function as compared to the conventional mean-field based neural
approach.

A particular class of binary CSP that has attracted recent attention
is {\em $K$-SAT} \shortcite{pap94,du97}; many combinatorial
optimization problems can be cast in $K$-SAT form.  We will
demonstrate in detail how to apply the information-based ANN approach,
to be referred to as {\em INN}, to $K$-SAT as a modified mean-field
annealing algorithm.

The method is evaluated by means of extensive numerical explorations
on suitable testbeds of random $K$-SAT instances. The resulting
performance shows a substantial improvement as compared to that of the
conventional ANN approach, and is comparable to that of a good dedicated
heuristic -- {\em Gsat+Walk} \shortcite{sel94,gu97}.

The real strength of the INN approach lies in its generality -- the
basic idea can easily be applied to arbitrary types of constraint
satisfaction problems, not necessarily binary.

\section{$K$-SAT}

A CSP amounts to determining whether a given set of simple constraints
over a set of discrete variables can be simultaneously fulfilled.

Most heuristic approaches to a CSP attempt to find a {\em solution},
i.e. an assignment of values to the variables consistent with the
constraints, and are hence {\em incomplete} in the sense that they
cannot prove unsatisfiability. If the heuristic succeeds in finding a
solution, satisfiability is proven; a failure, however, does not imply
unsatisfiability.

A commonly studied class of binary CSP is {\em $K$-SAT}.  A $K$-SAT
instance is defined as follows: For a set of $N$ Boolean variables
$x_i$, determine whether an assignment can be found such that a given
Boolean function $U$ evaluates to True, where $U$ has the form
\beq
	U = \(a_{11} \OR a_{12} \OR \dots a_{1K} \)
	\AND \(a_{21} \OR \dots a_{2K}\)
	\AND \dots
	\AND \(a_{M1} \OR \dots a_{MK}\) \; ,
\eeq
i.e. $U$ is the Boolean disjunction of $M$ {\em clauses}, indexed by
$m = 1\dots M$, each defined as the Boolean conjunction of $K$ simple
statements (literals) $a_{mk}, k=1\dots K$. Each literal represents
one of the elementary Boolean variables $x_i$ or its negation $\neg
x_i$.

For $K=2$ we have a 2-SAT problem; for $K=3$ a 3-SAT problem, etc. If
the clauses are not restricted to have equal length the problem is
referred to as a {\em satisfiability} problem ({\em SAT}).
There is a fundamental difference between $K$-SAT problems for
different values of $K$.  While a 2-SAT instance can be exactly solved
in a time polynomial in $N$, $K$-SAT with $K \geq 3$ is
NP-complete. Every $K$-SAT instance with $K>3$ can be transformed in
polynomial time into a 3-SAT instance \shortcite{pap94}.
In this paper we will focus on 3-SAT.

\section{Conventional ANN Approach}

\subsection{ANN Approach to CSP in General}

In order to apply the conventional mean-field based ANN approach as a
heuristic to a Boolean CSP problem, the latter is encoded in terms of
a non-negative cost function $H(\s)$ in terms of a set of $N$ binary
($\pm 1$) spin variables, $\s = \{s_i, \; i = 1,\dots,N\}$, such that
a solution corresponds to a combination of spin values that makes the
cost function vanish.

The cost function can be extended to continuous arguments in a unique
way, by demanding it to be a {\em multi-linear} polynomial in the
spins (i.e. containing no squared spins).  Assuming a multi-linear
cost function $H(\s)$, one considers mean-field variables (or {\em neurons}) 
$v_i \in[-1,1]$, approximating the thermal spin averages $\langle s_i \rangle$
in a Boltzmann distribution $P(\s) \propto exp(-H(\s)/T)$. They are
defined by the mean-field equations
\beqa
\label{mfv}
	v_i &=& \tanh(u_i/T)
\\
\label{mfu}
	u_i &=& -\frac{\partial H(\v)}{\partial v_i} \; ,
\eeqa
where $u_i$ is referred to as the {\em local field} for spin
$i$. Here, $T$ is an artificial temperature and $\v$ denotes the collection
of mean-field variables.

The equations (\ref{mfv},\ref{mfu}) can be seen as conditions for a
local minimum of the mean-field {\em free energy} $F(\v)$,
\beq
	F(\v) = H(\v) - T S(\v) \; ,
\eeq
where $S(\v)$ is the spin entropy,
\beq
	S(\v) = -\sum_i \frac{1+v_i}{2} \log\(\frac{1+v_i}{2}\)
	- \frac{1-v_i}{2}\log\(\frac{1-v_i}{2}\) \; .
\label{entropy}
\eeq

The conventional ANN algorithm consists in solving the mean-field
equations (\ref{mfv}, \ref{mfu}) iteratively, combined with annealing
in the temperature. A typical algorithm is described in figure
\ref{mfalgor}.
\begin{figure}[htb]
\begin{center}
\begin{large}
\fbox{\parbox{15cm}{
\begin{itemize}
\item Initiate the mean-field spins $v_i$ to random values close to
	zero, and $T$ to a high value.
\item Repeat the following (a sweep), until the mean-field variables have
	{\em saturated} (i.e. become close to $\pm 1$):
 \begin{itemize}
 \item For each spin, calculate its local field from (\ref{mfu}), and
	update the spin according to (\ref{mfv}).
 \item Decrease $T$ slightly (typically by a few percent).
 \end{itemize}
 \item Extract the resulting solution candidate, using $s_i = \sign(v_i)$.
\end{itemize}
}}
\end{large}
\end{center}
\caption{A mean-field annealing ANN algorithm.}
\label{mfalgor}
\end{figure}
%

\subsection{Application to $K$-SAT}

When applying the ANN approach to $K$-SAT the Boolean variables are
encoded using $\pm 1$-valued spin variables $s_i$, $i=1\dots N$, with
$s_i=+1$ representing True, and $s_i=-1$ False. In terms of the spins,
a suitable multi-linear cost function $H(\s)$ is given by the
following expression,
\beq
	H(\s) = \sum_{m=1}^M \prod_{i\in \Mm}
	\half \( 1-C_{mi} s_i \) \; ,
\label{H}
\eeq
where $\Mm$ denotes the set of spins involved in the $m$th clause.  $H(\s)$
evaluates to the number of broken clauses, and vanishes iff $\s$
represents a solution. The $M\times N$ matrix $C$ defines the $K$-SAT
instance: An element $C_{mi}$ equals $+1$ (or $-1$) if the $m$th
clause contains the $i$th Boolean variable as is (or negated);
otherwise $C_{mi}=0$.

The cost function (\ref{H}) defines a problem-specific set of
mean-field equations, (\ref{mfv},\ref{mfu}), in terms of mean-field
variables $v_i\in [-1,1]$. In the mean-field
annealing approach (figure \ref{mfalgor}), the temperature $T$ is
initiated at a high value, and then slowly decreased (annealing),
while a solution to (\ref{mfv},\ref{mfu}) is tracked iteratively.
At high temperatures there will be a stable fixed point with all
neurons close to zero, while at a low temperature they will approach
$\pm 1$ (the neurons have {\em saturated}) and an assignment can be extracted.

For the $K$-SAT cost function (\ref{H}) the local field $u_i$ in
(\ref{mfu}) is given by
\beq
	u_i = \sum_m \half C_{mi} \prod_{\stackrel{j\in \Mm}{j\neq i}}
	\half \(1-C_{mj} v_j\) \; ,
\eeq
which, due to the multi-linearity of $H$ does not depend on $v_i$;
this lack of self-coupling is beneficial for the stability of the
dynamics.

\section{Information-Based ANN Approach: INN}

\subsection{The Basic Idea}

For problems of the CSP type, we suggest an information-based neural
network approach, based on the idea of balance of information,
considering the variables as {\em sources} of information, and the
constraints as {\em consumers} thereof.

This suggests constructing an {\em objective function} (or free energy) $F$ of 
the general form
\beq
	F = \mbox{const.} \times \mbox{(information demand)}
	- \mbox{const.} \times \mbox{(available information)} \; ,
\eeq
that is to be minimized.  The meaning of the two terms can be made
precise in a mean-field-like setting, where a factorized artificial
Boltzmann distibution is assumed, with each Boolean variable having an
independent probability to be assigned the value True. We will give a
detailed derivation below for $K$-SAT.  Other problem types can be
treated in an analogous way.
We will refer to this type of approach as {\em INN}.

\subsection{INN Approach to $K$-SAT}

Here we describe in detail how to apply the general ideas above to the
specific case of $K$-SAT.
%

The average information resource residing in a spin is given by its
entropy,
\beq
	S(s_i)= -P_{s_i=1}\log P_{s_i=1} - P_{s_i=-1}\log P_{s_i=-1} \; ,
\eeq
where $P$ are probabilities. If the spin is completely random,
$P_{s_i=1}=P_{s_i=-1}=\half$ and $S(s_i)=log(2)$, representing an
unused resource of one bit of information. If the spin is set to a
definite value ($s_i=\pm 1$), no more information is available and
$S(s_i)=0$.

For a clause the interesting property is the expected amount of
information needed to satisfy it. For the $m$th clause, this can be
estimated as
\beq
	I_m = -\log P_m^{\mbox{\footnotesize sat}}
	= -\log \(1 - P_m^{\mbox{\footnotesize unsat}}\) \; ,
\eeq
in terms of the probability $P_m^{\mbox{\footnotesize sat}}$ for the
clause to be satisfied in a given probability distribution for the
spins.

Of the $2^K$ distinct states available to the $K$ spins appearing in
the clause, only one corresponds to the clause being
unsatisfied. Then, for a totally undetermined clause (all $K$ spins
having random values), we have $P_m^{\mbox{\footnotesize unsat}} =
2^{-K}$, yielding $I_m = -\log\(1-2^{-K}\)$.  For a definitely
satisfied clause, on the other hand, we must have
$P_m^{\mbox{\footnotesize unsat}} = 0$, giving $I_m = 0$.  Finally, a
broken clause corresponds to $P_m^{\mbox{\footnotesize unsat}} = 1$,
leading to $I_m \rightarrow \infty$.

Assuming a mean-field-like probability distribution, with each spin
obeying independent probabilities
\beq
	P_{s_i = \pm 1} = \frac{1 \pm v_i}{2} \; ,
\eeq
in terms of mean-field variables $v_i = \langle s_i \rangle \in
[-1,1]$, the probabilities used above for the clauses become
\beq
	P_m^{\mbox{\footnotesize unsat}} = \prod_{i \in \Mm }
	\half \(1-C_{mi} v_i\) \; .
\eeq
The unused spin information is given by the entropy $S$ of the spins (eq.
(\ref{entropy})) and the information $I$ needed by the clauses is
\beq
	I(\v) = \sum_{m=1}^M
	-\log \( 1 - \prod_{i \in \Mm } \half \(1-C_{mi} v_i\)\) \; .
\label{I}
\eeq

We now have the necessary prerequisites to define an information-based
free energy, which we choose as $F(\v) = I(\v) - T S(\v)$ (in analogy with 
ANN), which is to be minimized.
Demanding that $F$ have a local minimum with respect to the mean-field
variables yields equations similar to the mean-field equations
(\ref{mfv},\ref{mfu}), but with $H(\v)$ replaced by $I(\v)$:
\beq
	u_i = -\frac{\partial I}{\partial v_i} \; .
\label{uI}
\eeq
Note that for discrete arguments, $v_i=\pm 1$, the infomation demand
$I$ will be infinite for any non-solving assignment.

\subsection{Algorithmic Details}

Based on the analysis above, we propose an information-based
annealing algorithm similar to mean-field annealing, but with the
multi-linear cost function $H$ (\ref{H}) replaced by the clause
information $I$ (\ref{I}).

Note that the contribution $I_m$ to $I$ from a single clause $m$
is a simple function of the corresponding contribution $H_m$ to $H$,
\beq
	I_m = -\log \(1 - H_m\) \; .
\eeq
As a result, the effective cost function $I$ is not multilinear, and
measures have to be taken to ensure stability of the dynamics. The
resulting self-couplings can be avoided by instead of the derivative
in (\ref{uI}) using the difference,
\beq
	u_i = -\half \( \left. I \right|_{v_i=1}
	- \left. I \right|_{v_i=-1} \) \; ,
\label{uIdiff}
\eeq
which coincides with the derivative for a multilinear $I$
\shortcite{ohl93}.

The resulting INN annealing algorithm is summarized in figure
\ref{Ialgorithm}.
\begin{figure}[htb]
\begin{center}
\begin{large}
\fbox{\parbox{15cm}{
 \begin{enumerate}
 \item Choose a suitable high initial temperature $T$, such that the
      equilibrium neurons are close to zero.
 \item Do a sweep: Update all neurons according to
       (\ref{mfv},\ref{uIdiff}).
 \item Lower the temperature $T$ by a fixed factor $\mu$.
 \item If the stop-criteria are not met, repeat from 2.
 \item Extract a solution by means of $s_i=\sign(v_i)$.
 \end{enumerate}
 A typical $\mu$ value is 0.95 - 0.99, and suitable stop-criteria are that
 all neurons are either saturated ($|v_i| > 0.99$) or redundant ($|v_i| < 0.01$).
}}
\end{large}
\end{center}
\caption{The INN annealing algorithm for $K$-SAT.}
\label{Ialgorithm}
\end{figure}
At high temperatures, information is expensive, and the neurons stay
fuzzy, $v_i \approx 0$. As $T$ is decreased, information becomes
cheaper and the more useful neurons begin to saturate. As $T \to 0$, all
neurons are eventually forced to saturate, yielding a definite spin
state, $v_i \approx s_i = \pm 1$.

\section{Numerical Explorations}

\subsection{Testbeds}

For performance investigations, we have considered two distinct
testbeds. One consists of uniform random $K$-SAT problems with $N$
and $\alpha = M/N$ fixed (\shortcite{coo97}).  For every problem
instance, each of the $M$ clauses is independently generated by
chosing at random a set of $K$ distinct variables (among the $N$
available). Each selected variable is negated with probability
$\half$.

For this ensemble of problems, the fraction unsatifiable problems
increases with the parameter $\alpha$. In the thermodynamic limit
($N\rightarrow \infty$) there is a sharp satisfiability transition at
a $K$-dependent critical $\alpha$-value $\alpha_c^{(K)}$
\shortcite{hog96,mon99}. For problems where $\alpha < \alpha_c^{(K)}$
almost all generated problems are satisfiable and for $\alpha >
\alpha_c^{(K)}$ almost all are unsatisfiable. For 3-SAT, $\alpha_c
\approx 4.25$ \shortcite{coo97,mon99}.

We have used a set of $N$-values between 100 and 2000, and for each
$N$ a set of $\alpha$-values between 3.7 and 4.3.  For each $N$ and
$\alpha$, 200 problem instances are generated.

In addition, testbeds consisting purely of satisfiable instances are
useful to gauge the efficiency of a heuristic. Such a testbed can be
generated by filtering out unsatisfiable instances (using a {\em
complete} (exact) algorithm) from the uniform random distribution
described above.

For a second testbed, we have collected a set of instances of this
type from SATLIB\footnote{\tt
http://www.informatik.tu-darmstadt.de/AI/SATLIB}, consisting in
satisfiable random problems for different $N$ between 20 and 250, with
$\alpha$ fixed close to $\alpha_c$. For natural reasons, this testbed
does not include very large $N$.

\subsection{Comparison Algorithms}

To gauge the performance of the INN algorithm, we have in addition to
the conventional ANN algorithm also applied a state-of-the-art
dedicated heuristic to our testbeds. 
A wealth of algorithms has been tested on SAT problems. For a survey, see
e.g. \shortciteNP{gu97}. A local search method proven to be competitive is the
{\em gsat+walk} algorithm which we will use as a second reference algorithm.

Gsat+walk starts with a random assignment and then uses two types of
local moves to proceed. A local move consists in flipping the state of
a single variable between True/False. The first type of move is
greedy; the flip that increases the number of satisfied clauses
the most is chosen. The second type of move is a restricted
random walk move. A clause among those that are unsatisfied is chosen
at random, and then a randomly chosen variable in this clause is
flipped.

\subsection{Implementations details}

In order to have a fair comparison of performances, we have chosen the 
parameter values such, that the three algorithms use approximately equal CPU 
time for each problem size.

\subsubsection{ANN}

For ANN a preliminary initial temperature of 3.0 is used, which is
dynamically adjusted upwards until the neurons are close to zero 
($\sum_i v_i^2 < 0.1 N$), in order to ensure a start close to the high-$T$ 
fixed point.

The annealing rate is set to 0.99. At each temperature up to 10 sweeps
are allowed in order for the neurons to converge, as
signalled by the maximal change in value for a single neuron being less
than 0.01.
At every tenth temperature value, the cost function is evaluated using
the signs of the mean-field variables, $s_i = \mbox{sign}(v_i)$; if
this vanishes, a solution is found and the algorithm exits.
If no solution has been found when the temperature reaches a certain
lower bound (set to 0.1), the algorithm also exits; at that
temparature, most neurons typically will have stabilized close to
$\pm 1$ (or occasionally 0). Neurons that wind up at zero are those that are 
not needed at all or equally needed as $\pm 1$.

\subsubsection{INN}

For the INN approach, the same temperature parameters as in ANN are
used except for the low $T$ bound, which is set to 0.5. Because of the
divergent nature of the cost function $I$ (\ref{I}) and the local
field $u_i$ (\ref{uIdiff}), extra precaution has to be taken when
updating the neurons -- infinities appear when all the neurons in a
clause are $\pm 1$ with the wrong sign: $v_i = -C_{mi}$. When
calculating $u_i$, the infinite clause contributions are counted
separately. If the positive (negative) infinities are more (less)
numerous, $v_i$ is set to +1 (-1); otherwise, $v_i$ is randomly set to
$\pm 1$ if infinities exist but in equal numbers, else the finite part
of $u_i$ is used.

This introduces randomness in the low temperature region if a solution
has not been found; the algorithm then acquires a local search
behaviour increasing its ability to find a solution.
In this mode the neurons do not change smoothly and the maximum
number of updates per temperature sweep (set to 10) is frequently used, which
explains why INN needs more time than the conventional ANN for difficult 
problem instances.
Performance can be improved, at the cost of increasing the CPU time
used, with a slower annealing rate and/or a lower low-$T$
bound. Restarts of the algorithm also improves performance.

\subsubsection{gsat+walk}

The source code for gsat+walk can be found at SATLIB
\footnote{\tt
http://www.informatik.tu-darmstadt.de/AI/SATLIB}. We have attempted
to follow the recommendations in the enclosed documentation for
parameter settings.
The probability at each flip of choosing a greedy move instead of a
restricted random walk move is set to 0.5.
We have chosen to use a single run with $200 \times N$ flips per
problem, instead of several runs with less flips per try,
since this appears to improve overall performance.
Making several runs or using more flips per run will improve
performance at the cost of an increased CPU consumption.

\subsection{Results}

Here follow the results from our numerical investigations for the two
testbeds. All explorations have been made on a 600 MHz AMD Athlon computer 
running Linux.

\begin{figure}[htb]
\begin{center}
\mbox{\includegraphics{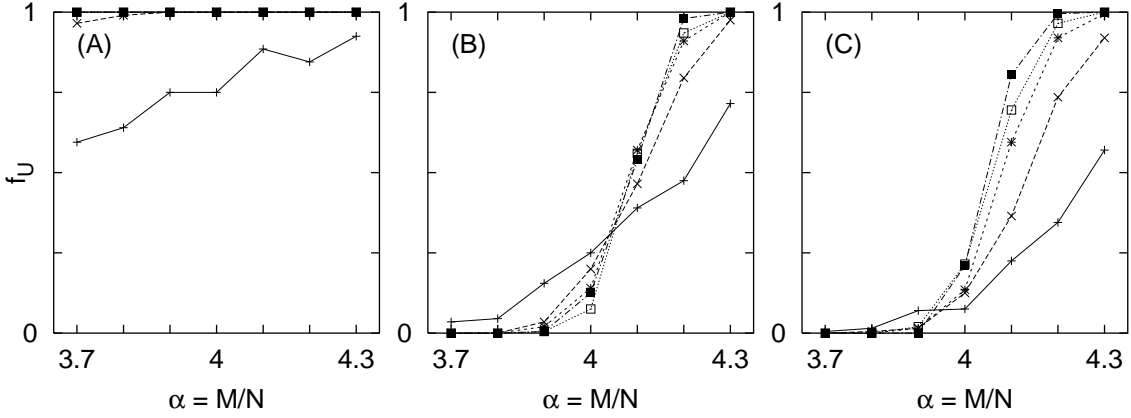}}
\end{center}
\caption{Fraction unsatisfied problems ($f_U$) versus $\alpha$ for ANN
 ({\bf A}), INN ({\bf B}) and gsat+walk ({\bf C}), for $N$ = 100 (+),
 500 ($\times$), 1000 ($\ast$), 1500 ({\tiny $\square$}) and 2000
 ({\tiny $\blacksquare$}). The fractions are calculated from 200 instances; the
 error in each point is less than 0.035.}
\label{fUres}
\end{figure}

\begin{figure}[htb]
\begin{center}
\mbox{\includegraphics{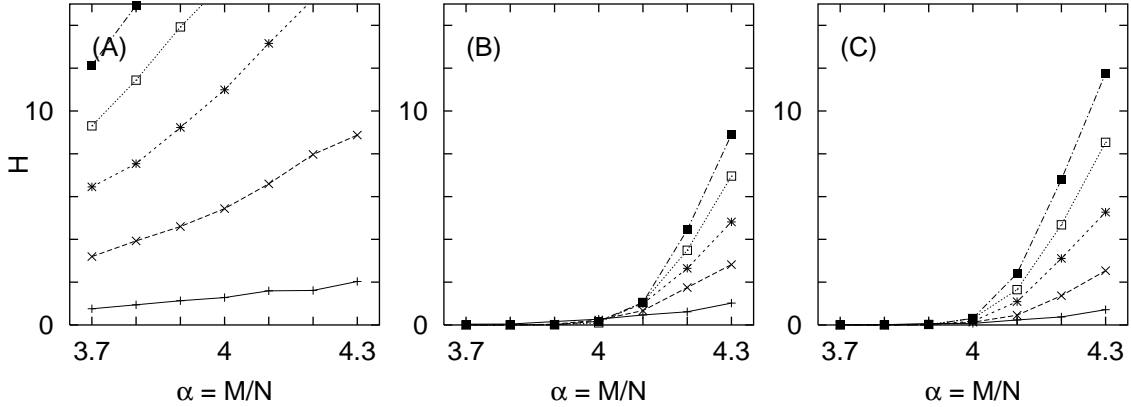}}
\end{center}
\caption{Number of unsatisfied clauses $H$ per instance versus
 $\alpha$, for ANN ({\bf A}), INN ({\bf B}) and gsat+walk ({\bf C}),
 for $N$ = 100 (+), 500 ($\times$), 1000 ($\ast$), 1500 ({\tiny $\square$}) and
 2000 ({\tiny $\blacksquare$}). Average over 200 instances.}
\label{Hres}
\end{figure}

\begin{figure}[htb]
\begin{center}
\mbox{\includegraphics{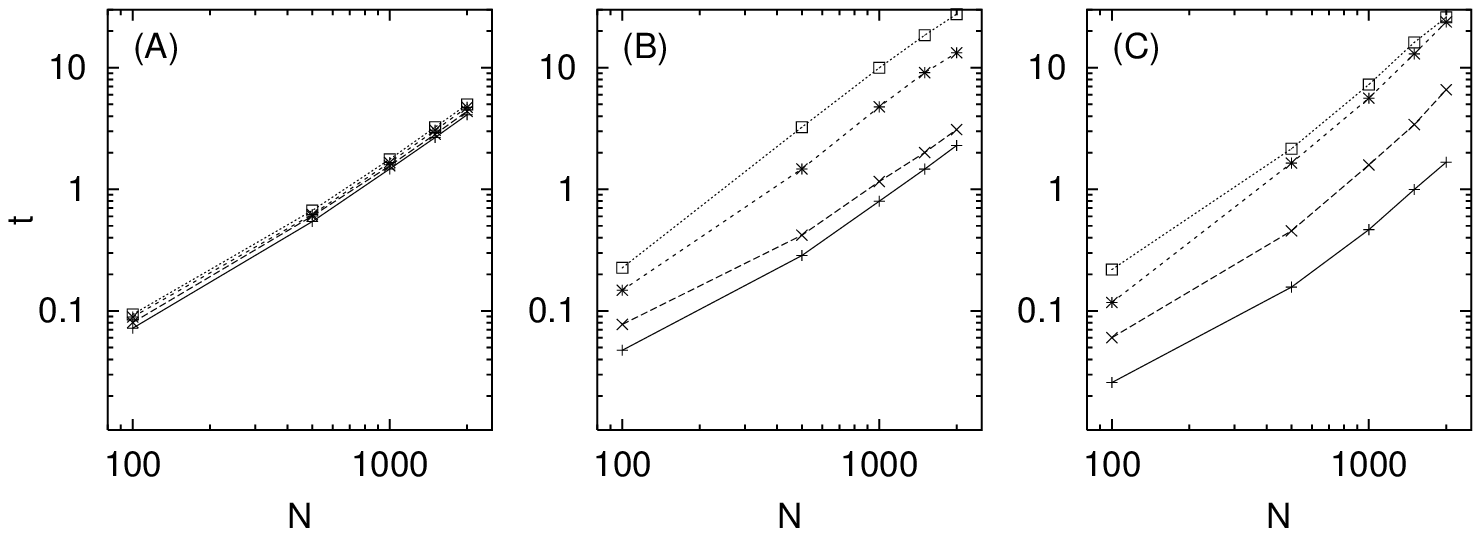}}
\end{center}
\caption{Log-log plot of used CPU-time $t$ (given in seconds) versus 
$N$, for ANN ({\bf A}), INN ({\bf B}) and gsat+walk ({\bf C}), for 
$\alpha$ = 3.7 (+), 3.9 ($\times$), 4.1 ($\ast$) and 4.3 ({\tiny $\square$}). 
$N$ ranges from 100 to 2000. Averaged over 200 instances.}
\label{TIMEres}
\end{figure}

The results from INN, ANN and Gsat+Walk for the uniform random testbed
are summarized in figures \ref{fUres}, \ref{Hres}, and
\ref{TIMEres}. 
  
In figure \ref{fUres} the fraction of the problems not satisfied by
the separate algorithms ($f_U$) is shown as a function of $\alpha$ for
different problem sizes $N$.  The three algorithms show different
transitions in $\alpha$ above which they fail to find solutions. For
INN and gsat+walk the transition appears slightly beneath the real
$\alpha_c$, while for ANN the transition is situated below
$\alpha=3.7$.
%
%

The average number of unsatisfied clauses per problem instance ($H$) is 
presented in figure \ref{Hres} for the three algorithms. $H$ is shown as a 
function of $\alpha$ for different $N$. This can be used as a performance 
measure also when an algorithm fails to find solutions
\footnote{Finding a maximal number of satisfied clauses for a SAT 
instance is referred to as MAXSAT \shortcite{pap94}.}.

The average CPU-time consumption ($t$) is shown in figure \ref{TIMEres} for 
all algorithms. The CPU-time is presented as a function of $N$ for different 
$\alpha$ in order to show how the algorithms scale with problem size.  

The results ($f_U$, $H$, $t$) for the solvable testbed for all three 
algorithms are summarized in table \ref{ufresult}.
%
%
\begin{table}[htb]
\begin{center}
\begin{tabular}{|rrr|rrr|rrr|rrr|} \hline
& &{\bf num} & \multicolumn{3}{|c|}{\bf ANN}
 & \multicolumn{3}{|c|}{\bf INN}
 & \multicolumn{3}{|c|}{\bf gsat+walk} \\
 $N$ & $M$ & {\bf inst.}
& $f_U$ & $H$ & $t$ & $f_U$ 
& $H$ & $t$ & $f_U$ & $H$ 
& $t$ \\ \hline
20  &91   &1000 &0.231 &0.248 &0.01 &0.076 &0.078 &0.01 &0.000 &0.000 &0.01 \\
50  &218  &1000 &0.607 &0.759 &0.04 &0.194 &0.208 &0.05 &0.008 &0.008 &0.02 \\
75  &325  &100  &0.84  &1.3   &0.07 &0.41  &0.44  &0.11 &0.05  &0.05  &0.05 \\
100 &430  &1000 &0.844 &1.485 &0.09 &0.315 &0.362 &0.13 &0.072 &0.074 &0.09 \\
125 &538  &100  &0.88  &1.72  &0.11 &0.39  &0.41  &0.18 &0.10  &0.10  &0.13 \\
150 &645  &100  &0.89  &2.07  &0.14 &0.34  &0.4   &0.23 &0.16  &0.17  &0.19 \\
175 &753  &100  &0.98  &2.6   &0.17 &0.51  &0.61  &0.39 &0.27  &0.28  &0.33 \\
200 &860  &100  &1     &3.06  &0.20 &0.6   &0.81  &0.52 &0.32  &0.34  &0.39 \\
225 &960  &100  &0.97  &3.15  &0.22 &0.52  &0.67  &0.51 &0.35  &0.37  &0.46 \\
250 &1075 &100  &0.99  &3.53  &0.25 &0.58  &0.77  &0.65 &0.39  &0.44  &0.53 \\ \hline
\end{tabular}
\end{center}
\caption{Results for the solvable 3-SAT problems close to
 $\alpha_c$. $f_U$ is the fraction of problems not satisfied by the
 algorithm, $H$ is the average number of unsatisfied clauses
 (\ref{H}) and $t$ is the average CPU-time used (given in seconds). The 
third column (num inst.) is the number of instances in the problem set.}
\label{ufresult}
\end{table}
%

\subsection{Discussion}

The first point to be made is the dramatic performance improvement in
INN as compared to ANN. This is partly due to the divergent nature of
the INN cost function $I$, leading to a progressively increased focus
on the neurons involved in the relatively few critical clauses on the
virge of becoming unsatisfied. This improves the revision capability
which is beneficial for the performance. The choice of randomizing
$v_i$ to $\pm 1$ (which appears very natural) in cases of balancing
infinities in $u_i$ contributes to this effect.

A performance comparison of INN and gsat+walk indicates that the
latter appears to have the upper hand for small $N$. For larger $N$ however, 
INN seems to be quite comparable to gsat+walk.
%

\section{Summary and Outlook}

We have presented a heuristic algorithm, INN, for binary
satisfiability problems. It is a modification of the conventional
mean-field based ANN annealing algorithm, and differs from this mainly
by a replacement of the usual multilinear cost function by one derived
from an information-theoretical argument.

This modification is shown empirically to dramatically enhance the
performance on a testbed of random $K$-SAT problem instances; the
resulting performance is for large problem sizes comparable
to that of a good dedicated heuristic, tailored to $K$-SAT.

An important advantage of the INN approach is its generality. The
basic philosophy -- the balance of information -- can be applied to a
host of different types of binary as well as non-binary problems; work
in this direction is in progress.
%

\section*{Acknowledgement}

Thanks are due to C. Peterson for inspiring discussions. 
This work was in part supported by the Swedish Foundation for Strategic 
Research.

\newpage

\bibliographystyle{chicago}
\bibliography{Ksat}

\end{document}